\documentclass[apjl]{emulateapj}
\usepackage{epsfig}
\usepackage{amsmath}
\usepackage{natbib}

\newbox\grsign \setbox\grsign=\hbox{$>$} \newdimen\grdimen \grdimen=\ht\grsign
\newbox\simlessbox \newbox\simgreatbox \newbox\simpropbox \newbox\wtildebox 
\setbox\simgreatbox=\hbox{\raise.5ex\hbox{$>$}\llap
    {\lower.5ex\hbox{$\sim$}}}\ht1=\grdimen\dp1=0pt
\setbox\simlessbox=\hbox{\raise.5ex\hbox{$<$}\llap
    {\lower.5ex\hbox{$\sim$}}}\ht2=\grdimen\dp2=0pt

\newcommand{\be}{\mbox{\begin{equation}}}
\newcommand{\ee}{\mbox{\end{equation}}}

\newcommand{\tdis}{\mbox{$t_{\rm dis}$}}

\newcommand{\Cref}{\mbox{$m_{\rm ref}$}}

\newcommand{\msun}{\mbox{M$_\odot$}}

\renewcommand{\d}{{\rm d}} 

\begin{document}

\submitted{Accepted for publication in ApJ Letters}

\title{On the interpretation of the globular cluster luminosity function}

\author{J.~M.~Diederik~Kruijssen\altaffilmark{1,2}}
\author{Simon~F.~Portegies~Zwart\altaffilmark{2,3}}
\altaffiltext{1}{Astronomical Institute, Utrecht University, PO Box 80000, 3508 TA Utrecht, The Netherlands; {\tt kruijssen@astro.uu.nl}}
\altaffiltext{2}{Leiden Observatory, Leiden University, PO Box 9513, 2300 RA Leiden, The Netherlands; {\tt spz@strw.leidenuniv.nl}}
\altaffiltext{3}{Astronomical Institute `Anton Pannekoek' and Section Computational Science, University of Amsterdam, 1098 SJ Amsterdam, The Netherlands}

\begin{abstract}
 The conversion of the globular cluster luminosity function (GCLF, $\d N/\d\log{L}$) to
 the globular cluster mass function (GCMF, $\d N/\d\log{M}$) is addressed. Dissolving
 globular clusters (GCs) become preferentially depleted in low-mass
 stars, which have a high mass-to-light ratio. This has been shown to
 result in a mass-to-light ratio ($M/L$) that increases with GC
 luminosity or mass, because more massive GCs have lost a smaller fraction of their stars
 than low-mass GCs. Using GC models, we study the
 influence of the luminosity dependency of $M/L$ on the inferred
 GCMF. The observed GCLF is consistent with a powerlaw or Schechter type GC initial mass
 function in combination with a cluster mass-dependent mass loss rate. Below the peak, the logarithmic slope of the GCMF is shallower than
 that of the GCLF (0.7 versus 1.0), whereas the peak mass is
 0.1---0.3~dex lower when accounting for the variability of $M/L$
 than in the case where a constant $M/L$ is adopted.
\end{abstract}

\keywords{
Galaxy: kinematics and dynamics --- 
(Galaxy:) globular clusters: general --- 
galaxies: kinematics and dynamics --- 
galaxies: star clusters ---
stellar dynamics}

\shortauthors{KRUIJSSEN \& PORTEGIES~ZWART}
\shorttitle{ON THE INTERPRETATION OF THE GLOBULAR CLUSTER LUMINOSITY FUNCTION}



\section{Introduction} \label{sec:intro}
\setcounter{footnote}{0}
The present-day globular cluster mass function (GCMF, $\d N/\d\log{M}$) is derived from the globular cluster luminosity function (GCLF, $\d N/\d\log{L}$) by assuming a constant mass-to-light ratio for all globular clusters \citep[e.g.,][]{fall01,vesperini03,jordan07,mclaughlin08}. The resulting GCMF is strongly depleted in low-mass globular clusters (GCs) with respect to the mass distribution of young star clusters, which is well-described by a powerlaw with index $-2$ in various environments down to a few 100~$\msun$. This has led to a number of {pioneering} studies explaining its shape by cluster evaporation at a cluster mass-independent mass loss rate \citep[equivalent to a disruption time $\tdis\propto M$, e.g.,][]{fall01,vesperini01} {acting on a powerlaw or \citet{schechter76} cluster initial mass function \citep[CIMF, e.g.,][]{harris94,mclaughlin96,elmegreen97,burkert00,gieles06b}. 

Although the observed peaked shape of the GCMF is reproduced in the above studies, the underlying assumptions are not entirely satisfactory because a cluster mass-dependent mass loss rate (equivalent to $\tdis\propto M^\gamma$ with $\gamma\sim 0.7$, see Eq.~\ref{eq:dmdt}) is found in theory \citep[e.g.,][]{baumgardt01,baumgardt03,gieles08} and observations \citep[e.g.,][]{lamers05,gieles08b,larsen09,gieles09}. This arises from the non-linear scaling of the disruption time with the half-mass relaxation time ($\tdis\propto t_{\rm rh}^{0.75}$), which is caused by the non-zero escape time of stars with velocities above the escape velocity from a tidally limited cluster \citep{fukushige00}.} {The physical effect of a lower $\gamma$ is that the dissolution rate of low-mass clusters is slowed down relative to higher cluster masses and higher $\gamma$.}

\begin{figure}
\resizebox{\hsize}{!}{\includegraphics{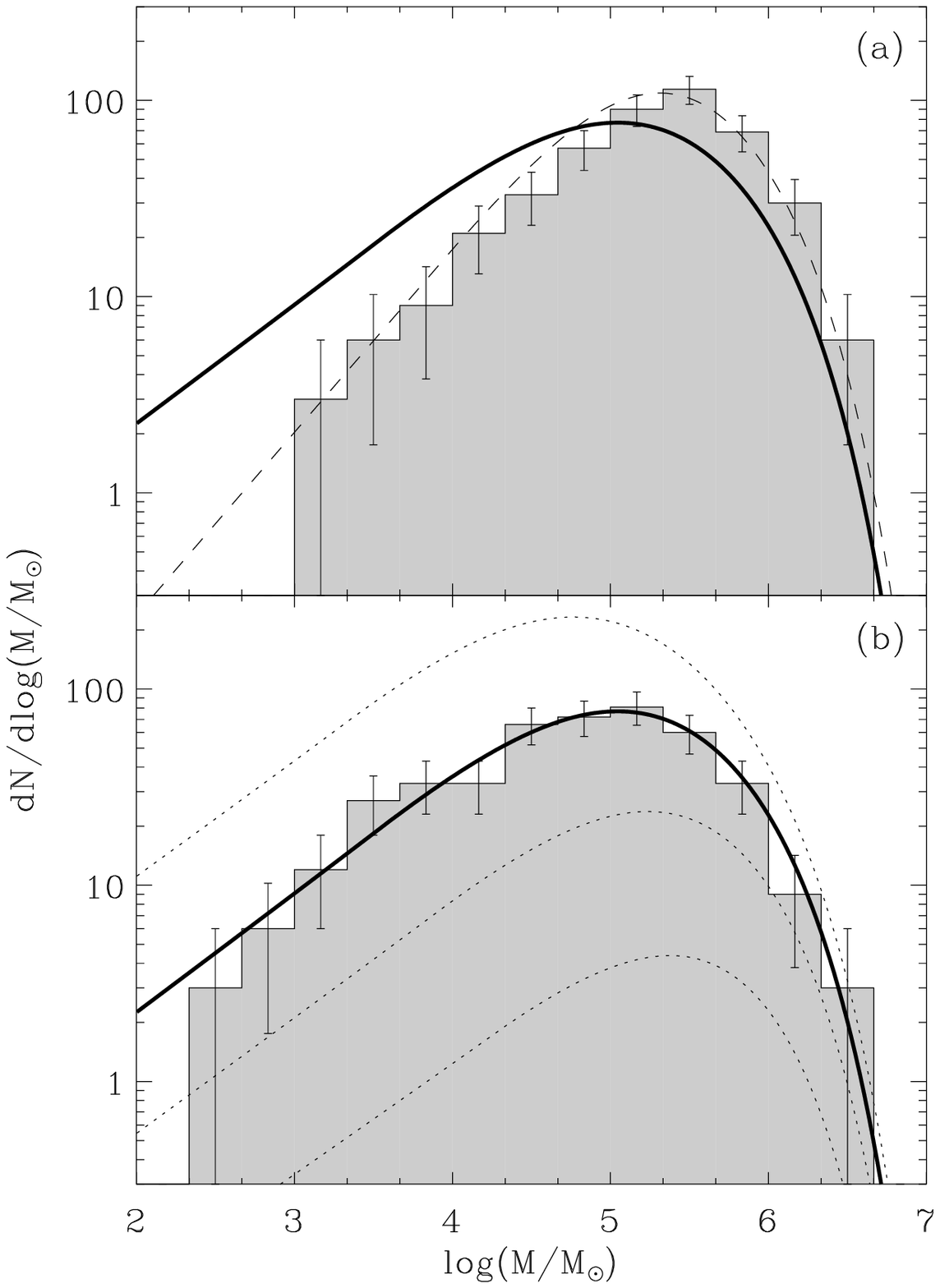}}
\caption[]{\label{fig:histcomp} The {inferred} GCMF of Galactic GCs \citep[histogram]{harris96}. {Panel (a)}: GCMF derived using $M/L_V=3$ \citep[as
   in][]{fall01}. Overplotted is our model MF {with a mass-dependent mass loss rate} (solid line, see
 Sect.~\ref{sec:simple}) adopting a dissolution timescale $t_0=1.3$~Myr. Using $\tdis = t_0 M^\gamma$
 \citep{lamers05}, for a $10^6~\msun$ GC {and $\gamma=0.7$} this corresponds to a
 disruption time of $\tdis= 21$~Gyr. The dashed line shows the
 model for a cluster mass-independent mass loss rate \citep[as
   in][]{fall01}. {Panel (b)}: GMCF derived from the GCLF using the luminosity dependent $M/L_V$
 (see Fig.~\ref{fig:mlcomp}). The solid curve is the same as above while the dotted curves represent models for (from bottom to top) $\log{(t_0/{\rm Myr})}=\log{1.3}+\{-0.5,-0.25,0.25\}$, {corresponding to $\tdis=7$---$37$~Gyr}. Error bars are 1$\sigma$ Poissonian.}
\end{figure}

The low-mass slope of a dissolution-dominated mass function like the GCMF is always equal to the {exponent $\gamma$} \citep{fall01,lamers05}. {A mass-dependent mass loss rate conflicts with the observations, as it yields a higher number of low-mass GCs compared to cluster mass-independent mass loss ($\gamma=1$).} The disagreement between cluster mass-dependent mass loss and the observed sparse population of
low-mass GCs is illustrated in Fig.~\ref{fig:histcomp}{(a)}. The slope of the modeled low-mass GCMF is
$\sim0.7$ for mass-dependent mass loss ($\gamma=0.7$), whereas for cluster mass-independent mass loss ($\gamma=1$) the slope is
$\sim1.0$, in agreement with the observations. The peak (or `turnover') masses also differ by
$\sim0.3$~dex. {These differences show that a lower mass loss rate for low-mass GCs ($\gamma=0.7$) yields a higher number of these relative to massive GCs than in the case of a constant mass loss rate ($\gamma=1$).}

Recent studies show that the mass-to-light ($M/L$) ratios of GCs are
not constant with luminosity or mass \citep{rejkuba07,kruijssen08b},
contrary to the assumption of a constant $M/L$ ratio in previous
studies. This agrees with an earlier analysis by
\citet{mandushev91}, who determined dynamical masses of Galactic GCs. These studies show that $M/L$ increases with mass
and luminosity because low-mass GCs are more strongly depleted in
low-mass stars. This variation of $M/L$ will affect the conversion of
the GCLF to a mass function. Specifically, the smaller $M/L$ ratios of low-mass GCs imply that
the masses of low-mass clusters are overestimated and consequently,
that the low-mass end of the GCMF would be shallower than presently
expected. The variability of $M/L$ could therefore strongly affect the
interpretation of the GCLF.

We show that the relation between the GCLF and the GCMF is affected by low-mass star depletion, which arises from two-body relaxation
\citep[e.g.,][]{meylan97}. In Sect.~\ref{sec:simple} we discuss the
influence of a luminosity-dependent mass-to-light ratio on the
inferred GCMF, and we model the GCLF in Sect.~\ref{sec:gclf},
alleviating the observationally expensive need for accurate $M/L$ ratios to
derive the GCMF to allow for a comparison with theory. {By including a mass-dependent mass loss rate and a variable $M/L$ ratio, our model provides an improvement to the \citet{fall01} model.}

\section{Implications of a luminosity-dependent $M/L$} \label{sec:simple}
We model the evolution of star clusters in order to quantify the
influence of the luminosity dependence of $M/L$ on the relation
between the GCLF and the GCMF.  Our model, called {\tt
 SPACE} \citep{kruijssen08}, includes mass loss by stellar evolution and by evaporation. The mass loss by
evaporation is parametrised with the simple relation \citep{lamers05}:
\begin{equation}
\label{eq:dmdt}
\left(\frac{\d M}{\d t}\right)_{\rm dis} = -\frac{M}{t_{\rm dis}} = -\frac{M^{1-\gamma}}{t_0}.
\end{equation}
Here $\gamma=0.7$ for clusters with a King parameter typical to
GCs of $W_0=7$ \citep{lamers09} and $t_0$ is the dissolution timescale
which depends on the environment. We illustrate the effect of a
variable $M/L$ ratio by adopting a unique value for $t_0$, which we assume
to be the same for all clusters (a realistic spread in $t_0$ is considered in Sect.~\ref{sec:gclf}). We subsequently convert the observed
LF of the sample of GCs to a MF by adopting the corresponding relation
between $L_V$ and $M/L_V$ that is computed with {\tt SPACE} {(see Fig.~\ref{fig:mlcomp})}. In
Fig.~\ref{fig:histcomp}{(b)} we show the resulting MF for the 146 GCs from
the \citet{harris96} catalogue\footnote{We adopt the 2003 edition of
 the data, which is available online at
 \texttt{http://www.physics.mcmaster.ca/\~{}harris/mwgc.dat}.}.
Overplotted are the model MFs with different values for $t_0$, {adopting a metallicity $Z=0.0004$, a \citet{kroupa01} stellar IMF and a Schechter CIMF with powerlaw index $-2$ and exponential truncation mass $M_*=2.5\times10^6~\msun$}. As expected from Eq.~\ref{eq:dmdt}, the slope of the MF is
independent of the dissolution timescale.

By comparing {panels (a) and (b) in Fig.~\ref{fig:histcomp}} we see
that the luminosity dependency of $M/L$ gives
rise to two effects: (1) the slope at the low-mass end of the {inferred}
GCMF drops to $\sim0.7$, which is the expected value for models with
cluster a mass-dependent cluster mass loss rate \citep{lamers05}, and (2) the peak in the MF (the so-called turnover mass) shifts to a lower mass with $\sim0.3$~dex. About
half this shift is due to the already high value of $M/L=3$ adopted by
\citet{fall01}. The slope of the GCMF at the low-mass end is different
from the slope of the GCLF, and therefore also different from the
GCMF slope ($\sim 1$) that would be inferred from the GCLF when
using a constant $M/L$ ratio.

\section{Models of the Galactic GC system} \label{sec:gclf}
In our above analysis we have assumed a single dissolution timescale for the entire GC system. In reality there is a range of timescales on which the GCs dissolve. We now consider a more detailed Monte Carlo model of the Galactic GC system in which the dependency of the dynamical evolution of GCs on their orbits is included. Our aim is to directly model the GCLF, rather than to obtain it by converting the GCMF.

The initial positions of the GCs with respect to the Milky Way are taken from the powerlaw-like density profile \citep[see e.g.,][Eq.~26]{fall01} that arises from the isothermal sphere, with an outward increase of the velocity anisotropy \citep{eddington15}. Our choice of parameters for the kinematic model are (1) an initial anisotropy radius $R_{\rm A}=1$~kpc, (2) a circular velocity of the gravitational potential $V_{\rm c}=220~{\rm km~s}^{-1}$, and (3) $(V_{\rm c}/v)^2=3.5$, which determines the slope of the density profile \citep{fall01}, with $v$ denoting the radial velocity dispersion. The initial velocities of the GCs are assigned according to the corresponding velocity ellipsoid \citep[Eq.~4]{aguilar88}, including a systemic rotation of $V_{\rm rot}=60~{\rm km~s}^{-1}$. We do not claim that this is the correct kinematic model for the Milky Way, but we consider it an appropriate ansatz. {The resulting dissolution timescales agree with the range that is expected from observations. For less anisotropy, the mean dissolution timescale of surviving clusters would be longer.} The initial cluster masses are drawn from a \citet{schechter76} function with index $-2$ and exponential truncation mass $M_*=3\times10^6~\msun$ {\citep[also see][]{jordan07,harris09}}\footnote{This number is slightly larger than in Sect.~\ref{sec:simple}, because the spread in dissolution timescales implies that surviving massive GCs on average have a smaller dissolution timescale than surviving low-mass GCs. We correct for the resulting deficiency of massive GCs by increasing $M_\star$.}. We sample the metallicities from their observed distribution in the \citet{harris96} catalogue. 

The GC orbits are integrated in the Galactic potential from \citet{johnston95}, {consisting of a bulge, disc and halo}. We adopt a 4$^{\rm th}$-order Runge-Kutta integration scheme with a variable timestep, in which the angular momentum and energy are conserved within $10^{-5}$ during each timestep. To compute the evolution of a GC with a given initial mass and metallicity, we derive its instantaneous dissolution timescale from the orbital parameters. Tidal evaporation due to two-body relaxation and disc shocks are the main dissolution mechanisms \citep{chernoff86}. Following \citet{baumgardt03} for the dissolution timescale due to two-body relaxation we write:
\begin{equation}
 \label{eq:t0evap}
 t_{0,{\rm evap}}=10.7~{\rm Myr}~\left(\frac{R_{\rm a}}{8.5~{\rm kpc}}\right)\left(\frac{V_{\rm c,a}}{220~{\rm km~s^{-1}}}\right)^{-1}(1-e) ,
\end{equation}
where $R_{\rm a}$ is the apogalactic radius of the cluster orbit, $V_{\rm c,a}$ is the circular velocity of the gravitational potential at $R_{\rm a}$, and $e$ is the orbital eccentricity. The dissolution timescale due to disc shocks is expressed as \citep[Kruijssen et al., in prep.]{gnedin97}:
\begin{equation}
 \label{eq:t0disc}
 t_{0,{\rm disc}}=7.35~{\rm Myr}~\left(\frac{V_{\rm z,5}}{g_{\rm m,-10}}\right)^2 P_{\rm 2}A^{-1}_{\rm w}(x) ,
\end{equation}
where $V_{\rm z,5}$ is the velocity in the $z$-direction during disc crossing at $z=0$ in units of 10$^5$~m~s$^{-1}$, $P_{\rm 2}$ is the (radial) orbital period in units of 10$^2$~Myr, $g_{\rm m,-10}$ is the orbital maximum of the acceleration due to the disc $-\partial\Phi_{\rm disc}/\partial z$ in units of 10$^{-10}$~m~s$^{-2}$, and $A_{\rm w}(x)$ is the \citet{weinberg94a,weinberg94b,weinberg94c} adiabatic correction\footnote{The parameter $x$ implicitly depends on the GC mass \citep[e.g.,][]{gnedin97}. We adopt 0.6 times the initial GC mass, in agreement with the average mass loss per Hubble time from \citet{kruijssen09}.} \citep[see also][]{gnedin97}. For mathematical simplicity, we assume a very weak mass-radius relation\footnote{Compared to adopting a constant GC radius, this assumption effects a $\sim0.45$~dex scatter of $t_{0,{\rm disc}}$. Because $\log{(t_{0,{\rm disc}}/t_{0,{\rm tot}})}>0.75$ for all surviving GCs, this does not affect our results.} of $r_{\rm h}\propto M^{0.1}$ \citep{larsen04b} in the derivation of Eq.~\ref{eq:t0disc}.
\begin{figure}
\resizebox{\hsize}{!}{\includegraphics{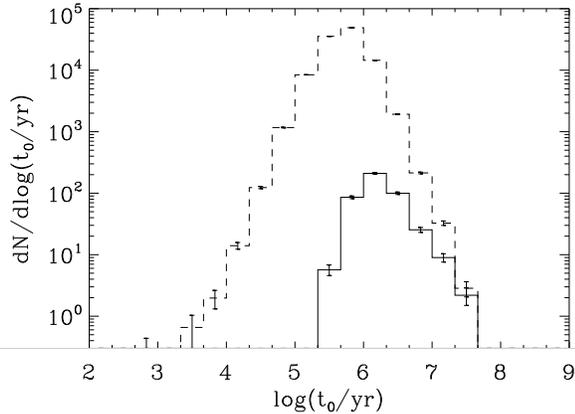}}
\caption[]{\label{fig:t0}
     Histograms of the initial (dashed) and present-day (solid) distributions of dissolution timescales $t_{0,{\rm tot}}$.
         }
\end{figure}

Over a timespan of 12~Gyr, the dissolution timescales due to tidal evaporation and disc shocks are computed for every orbital revolution, measured between subsequent passages of the apogalacticon. The dissolution timescale that describes the mass loss rate (see Eq.~\ref{eq:dmdt}) due to both effects is determined by adding the averaged inverses of both timescales:
\begin{equation}
 \label{eq:t0tot}
 \frac{1}{t_{0,{\rm tot}}}=\frac{1}{t_{0,{\rm evap}}}+\frac{1}{t_{0,{\rm disc}}} .
\end{equation}
The resulting initial and present-day distributions of $t_{0,{\rm tot}}$ are shown in Fig.~\ref{fig:t0}. GCs with short dissolution timescales are easily destroyed, leading to the depletion of the quickly-dissolving end of the distribution (at low values of $t_0$). The surviving GCs have dissolution timescales that are in excellent agreement with other studies \citep{kruijssen08b,kruijssen09}. {Although their mean galactocentric radius is a factor two smaller than that of the observed Galactic GC system, the slopes of both density profiles are comparable.}

\begin{figure}
\resizebox{\hsize}{!}{\includegraphics{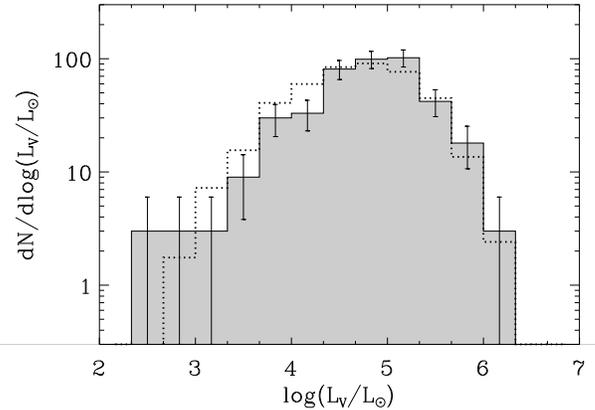}}
\caption[]{\label{fig:gclf}
      Histograms of the observed (filled) and modeled (dotted) GCLFs.
         }
\end{figure}
The evolution of GC mass and photometry is computed with {\tt SPACE}, using the setup discussed in Sect.~\ref{sec:simple}. In total {507079} GCs are generated with initial masses $M\geq5\times10^3~\msun$, of which {2000} survive until $t=12$~Gyr. The present-day mass and luminosity functions are scaled to match the observed numbers, of which the scale factor can be used to derive properties of the initial Galactic GC system (see below). The computed $V$-band GCLF is compared to the observed distribution in Fig.~\ref{fig:gclf}. The distributions are in satisfactory agreement, with a KS-test $p$-value of {0.02}. At low luminosities there is a slight discrepancy, which could be caused by incompleteness {due to obscuration by the Galactic bulge} (Gieles et al., in prep.). 

\begin{figure}
\resizebox{\hsize}{!}{\includegraphics{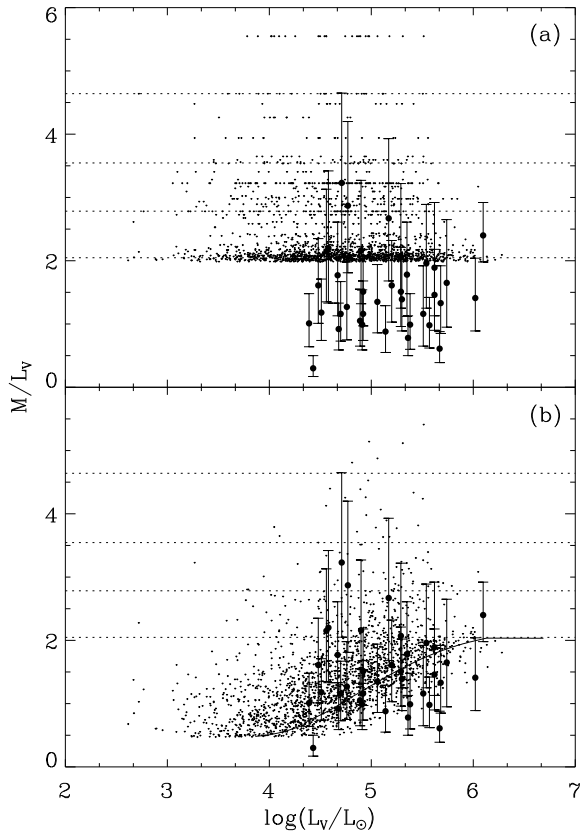}}
\caption[]{\label{fig:mlcomp}
     Comparison of observed $M/L_V$ ratios of Galactic GCs \citep[thick dots,][]{mclaughlin05} with our modeled $M/L_V$ ratios (thin dots). Error bars are 1$\sigma$. Dotted horizontal lines denote the constant $M/L_V$ ratios that are expected if low-mass star depletion is neglected (from bottom to top $Z=\{0.0004,0.004,0.008,0.02\}$). Panel (a): thin dots represent modeled $M/L_V$ ratios without low-mass star depletion. Panel (b): thin dots denote modeled $M/L_V$ ratios including low-mass star depletion. The solid line represents the relation between $M/L_V$ and $L_V$ that was adopted in the simple model of Fig.~\ref{fig:histcomp}(b).}
\end{figure}

{The $M/L_V$ ratios of our modeled GCs are compared to observations from \citet{mclaughlin05} in Fig.~\ref{fig:mlcomp}. If low-mass star depletion is neglected (panel (a)), the $M/L_V$ ratios of the models are completely set by their metallicities and they agree poorly with the observations. When including low-mass star depletion (panel (b)), the modeled $M/L_V$ ratios are affected by dynamical evolution and are in good agreement with the observations. The same approach can be used to explain the observations of Cen A, M31 and the LMC compiled by \citet{rejkuba07}, which gives results that are consistent with our analysis in Fig.~\ref{fig:mlcomp}. 

The \citet{mclaughlin05} sample is not representative of the entire Galactic GC population, as it lacks GCs that are much fainter than the turnover and represents central rather than global $M/L_V$ ratios for certain GCs, only allowing for a first-order comparison \citep[for a discussion, see][]{kruijssen09}. The observed slopes of the low-mass stellar mass functions of 20 GCs from \citet{demarchi07} provide an independent check. Their compilation exhibits a clear trend of mass function slope with GC luminosity. Splitting their sample at about the turnover luminosity (${\rm log}(L_V/L_\odot)=5.1$), for a mass function $n\propto m^{-\alpha}$ the mean slopes in the stellar mass range $m=0.3$---$0.8~\msun$ are $\alpha_{\rm bright}=1.42\pm0.10$ and $\alpha_{\rm faint}=0.56\pm0.07$ for the bright and faint GCs, respectively. Faint GCs are indeed more depleted in low-mass stars than bright GCs, substantiating our model results.}

\begin{figure}
\resizebox{\hsize}{!}{\includegraphics{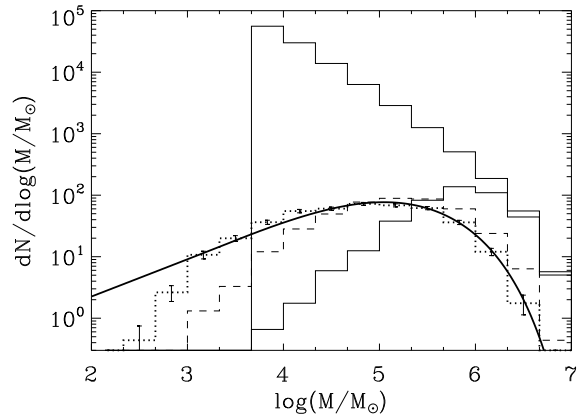}}
\caption[]{\label{fig:gcmf}
    Histograms of the mass distributions of the modeled GCs. Represented are the CIMF (upper solid) and the present-day GCMF (dotted, with 1$\sigma$ Poissonian error bars). The GCMF for a single dissolution timescale (Fig.~\ref{fig:histcomp}{(b)}) is represented by the continuous solid curve. The dashed line gives the GCMF that would be obtained from Fig.~\ref{fig:gclf} if a constant $M/L$ ratio were adopted. The initial mass distribution of the surviving GCs is given by the lower solid line. 
         }
\end{figure}

{The initial properties of the Galactic GC system are obtained by scaling the present-day number of modeled GCs to the observed number and applying the same scale factor to the CIMF.} In Fig.~\ref{fig:gcmf} we show the CIMF, the modeled GCMF, the GCMF that would be obtained from Fig.~\ref{fig:gclf} if a constant $M/L$ ratio were adopted, and the initial mass distribution of the surviving GCs. The modeled GCMF for a single dissolution timescale from Fig.~\ref{fig:histcomp}{(b)} is overplotted for comparison, illustrating its acceptable agreement with our detailed model. {The disagreement for GC masses $<10^3~\msun$ is due to the use of logarithmic timesteps in our models, causing some GCs to lose their last few $100~\msun$ within a single timestep at large ages.} For a lower mass limit of the CIMF of $5\times10^3$ ($10^2$)$~\msun$, we find a surviving GC number fraction of {3.9} $(0.1)\times10^{-3}$, with an initial total mass of about 1.1 $(1.8)\times10^9~\msun$ and a present-day mass of ${2.8\times10^7~\msun}$. If the stellar halo \citep[$\sim10^9~\msun$,][]{bell08} is constituted by disrupted GCs and coeval stars {\citep[in spite of chemical analyses, e.g.,][]{gratton00}}, our comparable initial total GC mass implies that either nearly all star formation occurred in clusters at the epoch of GC formation, or that most of these stars now constitute the Galactic bulge.

\section{Discussion} \label{sec:disc}
We have shown that the interpretation of the GCLF as a {one-to-one} representation of the GCMF is incorrect. This follows from the $M/L$ ratio decrease due to the low-mass star depletion that arises from two-body relaxation. There is no equivalence of the luminosity function and the mass function as both have intrinsically different low-mass slopes ($\sim1$ and $\sim0.7$, respectively). In addition, the turnover mass is overestimated by 0.1---0.3~dex if a one-to-one conversion from GCLF to GCMF is applied, depending on the adopted $M/L$ ratio. We have shown that the present-day GCLF and GCMF arise from a cluster mass-dependent mass loss rate ($\tdis\propto M^{0.7}$ and $\tdis\propto t_{\rm rh}^{0.75}$), starting from a Schechter-type CIMF. {Therefore, neither is consistent with a cluster mass-independent mass loss rate \citep[e.g.,][]{fall01}.} The GCMF that is computed using a spread in dissolution timescale $t_0$ only marginally differs from that for a single, mean value of $t_0$.

The low-mass slope of a dissolution-dominated mass function like the GCMF is equal to $\gamma$ (see Eq.~\ref{eq:dmdt}), independent of the CIMF \citep{fall01,lamers05}. For cluster mass-dependent mass loss ($\gamma=0.7$) the {GCMF that is inferred from the GCLF is accurately matched by the models} (see Fig.~\ref{fig:histcomp}{(b)}). To verify whether this perhaps holds for all values of $\gamma$, we have also considered cluster mass-independent mass loss \citep[$\gamma=1$,][]{fall01} and found that the luminosity dependency of $M/L$ (see Fig.~\ref{fig:mlcomp}) is steepened compared to cluster mass-dependent mass loss. {The conversion of the observed GCLF to a GCMF then gives an inferred GCMF slope} that is even lower ($\sim0.6$), in bad agreement with the expected ($\sim1$) value. We conclude that the match between the models and the observations only exists for values of $\gamma\approx 0.7$. Of course, the precise description of mass loss does not affect the fundamental principle of low-mass star depletion due to two-body relaxation. The luminosity dependence of $M/L$ flattens the {inferred} low-mass GCMF in any scenario.

We have not yet {considered the radial variation of the turnover luminosity $L_{\rm TO}$}, which has been shown to be independent of galactocentric radius in M87 \citep{vesperini03}. {Our} prescription for dynamical evolution in Sect.~\ref{sec:gclf} yields a higher turnover luminosity near the galactic centre than at large distances. However, our method is aimed at investigating the influence of a representative spread in dissolution timescales on our results, rather than making an exact model of the Galactic GC system. It should be emphasised that the difference between the GCLF and the GCMF persists, even though it remains to be explained why $L_{\rm TO}$ appears to be constant. It could be that the outer GCs dissolve more rapidly than expected. Potential explanations could be that GCs on wide orbits originate from {accreted dwarf galaxies} \citep{prieto08}, or a dissolution mechanism that has not yet been included \citep[see also][]{kruijssen09}, like the dynamical implications of white dwarf kicks \citep{fregeau09}, stellar evolution \citep{vesperini03b,vesperini09}, or gas expulsion \citep{baumgardt07}.


{The results of this paper do not only apply to the Milky Way, but also to other galaxies.} We see that the properties of the {inferred} GCMF are affected by the
mass and luminosity dependence of $M/L$ that ensues from low-mass star depletion. It is advised for observational and theoretical studies to be cautious when comparing GCLFs and GCMFs. At present an observed GCMF cannot be accurately obtained, because for
most observed GCs only photometric masses are determined (for which by
definition a constant $M/L$ ratio is assumed) instead of dynamical
masses. Considering the intrinsically
different shapes of the GCLF and GCMF, the presently most feasible way
of comparing theory and observations would be if models of GC systems are aimed at explaining the GCLF rather than the
mass distribution.

\begin{acknowledgements}
{We thank the anonymous referee for constructive comments that improved the manuscript.} We acknowledge Dana Casetti-Dinescu, Mike Fall and Dean McLaughlin for stimulating discussions, and Mark Gieles for comments on the manuscript. JMDK is grateful to Sophie Goldhagen and Henny Lamers for support, advice and comments on the manuscript. The Kavli Institute for Theoretical Physics in Santa Barbara is acknowledged for their hospitality and hosting an excellent globular cluster workshop and conference. This research is supported by the Netherlands Advanced School for Astronomy (NOVA), the Leids Kerkhoven-Bosscha Fonds and the Netherlands Organisation for ScientiÞc Research (NWO), grant numbers 021.001.038 and 643.200.503.
\end{acknowledgements}

\end{document}